\documentclass[12pt,preprint]{aastex}

\begin{document}

\title{Observations of Galactic and Extra-galactic Sources From the BOOMERANG and SEST Telescopes}

\author{K. Coble\altaffilmark{1,2},
P.A.R. Ade\altaffilmark{3},
J.J. Bock\altaffilmark{4}, 
J.R. Bond\altaffilmark{5}, 
J. Borrill\altaffilmark{6}, 
A. Boscaleri\altaffilmark{7}, 
C. R. Contaldi\altaffilmark{5}, 
B.P. Crill\altaffilmark{8}, 
P. de Bernardis\altaffilmark{9}, 
P. Farese\altaffilmark{10}, 
K. Ganga\altaffilmark{11}, 
M. Giacometti\altaffilmark{9}, 
E. Hivon\altaffilmark{11}, 
V.V. Hristov\altaffilmark{11}, 
A. Iacoangeli\altaffilmark{9}, 
A.H. Jaffe\altaffilmark{12},
W. C. Jones\altaffilmark{11}, 
A.E. Lange\altaffilmark{11}, 
L. Martinis\altaffilmark{13}, 
S. Masi\altaffilmark{9}, 
P. Mason\altaffilmark{11}, 
P.D. Mauskopf\altaffilmark{14}, 
A. Melchiorri\altaffilmark{9}, 
T. Montroy\altaffilmark{15}, 
C.B. Netterfield\altaffilmark{16}, 
L. Nyman\altaffilmark{17,18}
E. Pascale\altaffilmark{7}, 
F. Piacentini\altaffilmark{9}, 
D. Pogosyan\altaffilmark{5},
G. Polenta\altaffilmark{9}, 
F. Pongetti\altaffilmark{9}, 
S. Prunet\altaffilmark{5}, 
G. Romeo\altaffilmark{19}, 
J.E. Ruhl\altaffilmark{20}, 
F. Scaramuzzi\altaffilmark{13}
}

\altaffiltext{1}{Dept. of Astronomy and Astrophysics, University of Chicago, 5640 South Ellis Ave., Chicago, IL 60615, USA, coble@hyde.uchicago.edu}
\altaffiltext{2}{Adler Planetarium and Astronomy Museum, 1300 South Lake Shore Drive, Chicago, IL 60605, USA}
\altaffiltext{3}{Queen Mary and Westfield College, London, UK }
\altaffiltext{4}{Jet Propulsion Laboratory, Pasadena, CA, USA }
\altaffiltext{5}{Canadian Institute for Theoretical Astrophysics, University of Toronto, Canada }
\altaffiltext{6}{National Energy Research Scientific Computing Center, LBNL, Berkeley, CA, USA }
\altaffiltext{7}{IROE-CNR, Firenze, Italy }
\altaffiltext{8}{Physics Dept., California State University, Dominguez Hills, Carson, CA, USA}
\altaffiltext{9}{Dipartimento di Fisica, Universita' La Sapienza, Roma, Italy}
\altaffiltext{10}{Princeton University, Princeton, NJ, USA}
\altaffiltext{11}{California Institute of Technology, Pasadena, CA, USA }
\altaffiltext{12}{Center for Particle Astrophysics, University of California, Berkeley,CA, USA }
\altaffiltext{13}{ENEA Centro Ricerche di Frascati, Via E. Fermi 45, 00044, Frascati, Italy}
\altaffiltext{14}{Dept. of Physics and Astronomy, Cardiff University, Cardiff, Wales, UK }
\altaffiltext{15}{Dept. of Physics, University of California, Santa Barbara, CA, USA}
\altaffiltext{16}{Depts. of Physics and Astronomy, University of Toronto, Canada }
\altaffiltext{17}{Swedish-ESO Submillimetre Telescope (SEST), European Southern Observatory, Casilla 19001, Santiago, CHILE}
\altaffiltext{18}{Onsala Space Observatory, SE-439 92 Onsala, Sweden}
\altaffiltext{19}{Istituto Nazionale di Geofisica, Roma,~Italy }
\altaffiltext{20}{Physics Dept., Case Western Reserve University, Cleveland, OH, USA }


\begin{abstract}
We present millimeter-wave observations of three
extra-galactic and six Galactic sources in 
the Southern sky. Observations were made at 90, 150, 240 and 400
GHz with resolutions of 18, 10, 14 and 12 arcmin respectively 
during the 1998 Antarctic long duration balloon flight of BOOMERANG.
Observations were also made with the SEST telescope, at 90 and 150 GHz with 
resolutions of 57 and 35 arcsec respectively. These 
observations can be used for calibrations of Cosmic 
Microwave Background (CMB) experiments as well as an 
understanding of the physical processes of the sources.
\end{abstract}



\section{Introduction}

Galactic millimeter-wave radiation arises from a combination
of synchrotron, free-free and dust emission. Compact millimeter-wave
sources in the Galaxy can potentially be used to calibrate
observations of the Cosmic Microwave Background (CMB).  Southern Galactic
sources are particularly useful for CMB observations that take place in
the Antarctic, where planets are observed only at low elevation.  Previous
millimeter-wave observations of Southern Galactic sources have
been made by Arnold et al. (1978), Cheung et al. (1980), Cox et al.
(1995), Ruhl et al. (1995), and Puchalla et al. (2002). For reviews of
radio observations of a subset of our Galactic sources, see e.g. Brooks
et al. (1998) and De Pree et al. (1999). Extra-galactic point
sources are expected to be an important foreground contaminant of CMB
observations (Toffolatti et al 1999, Sokasian et al. 2001).

BOOMERANG is a millimeter-wave telescope and receiver designed to
measure anisotropy in the Cosmic Microwave Background (CMB) from the Long
Duration Balloon (LDB) platform. BOOMERANG launched
from McMurdo station, Antarctica on December 29, 1998
for a 10.5 day flight. The instrument is described in detail
in Crill et al. (2002). CMB angular power spectrum results can be found in
de Bernardis et al. (2000), Netterfield et al. (2002) and Ruhl et al. (2002).
Diffuse dust is discussed in Masi et al. (2001). In addition to
observing the CMB, BOOMERANG observed three extra-galactic point sources
and six Galactic compact HII regions.  The flux measurement of
these sources are calibrated with the CMB dipole. Follow-up observations
of the three extra-galactic sources and two of the Galactic sources
were made using the SEST telescope at La Silla Observatory in
Chile, January 8-10, 2000.  A list of sources and
coordinates are given in Table \ref{tbl1}.

\placetable{tbl1}


\section{BOOMERANG Observations and Results}

The BOOMERANG telescope features a 1.2m primary mirror and
a cryogenic bolometric array receiver. There are two 90 GHz
detectors, both with a FWHM of 18$\pm$1 arcmin, six 150 GHz
detectors, with FWHMs of 9.2$\pm$0.5, 9.2$\pm$0.5, 9.7$\pm$0.5,
9.4$\pm$0.5, 9.9$\pm$0.5, and 9.5$\pm$0.5 arcmin, three 240 GHz
detectors with a FWHM of 14.1$\pm$1 arcmin, and four 400 GHz
detectors with a FWHM of 12.1$\pm$1 arcmin. The main
lobes of the beams from the 90, 150, and 400 GHz channels are
well-characterized by single Gaussians. The 240 GHz beams have
a supressed central region, accurately modeled by the superposition
of one Gaussian to fit the full beam plus an inverted Gaussian
that improves the fit in the central region.
Long term random pointing drift does not add significantly to the beam
size for the Galactic sources, since each of those observations were made
on a short timescale. The extra-galactic sources however,
are affected by long-term random pointing drifts, because
those observations were made throughout
the flight. A pointing uncertainty of 2.5 arcmin rms
adds in quadrature to the effective beam size and adds 1.4
arcmin rms to the beam uncertainty for the extra-galactic sources.
See Netterfield et al. (2002) for a more
detailed discussion of pointing uncertainty.
The 90, 150, and 240 GHz channels are calibrated
using the CMB dipole, with an uncertainty of 10\% to diffuse
beam-filling sources.
The 400 GHz channels are calibrated using interstellar dust
and degree-scale CMB anisotropy, with an uncertainty of 30\%
(Crill et al. 2002).

Observations with BOOMERANG are made by scanning the telescope at an
angular velocity of 2 degrees/sec in azimuth for the first half of the flight
and 1 degree/sec in azimuth for the second half of the flight.
Interspersed between CMB observations are observations
of the Galactic plane.
The Galactic sources were observed at the same scan speeds
as the CMB observations, but with shorter
scans (10 degrees peak-to-peak in azimuth for the source
scans vs. 60 degrees peak-to-peak for the CMB scans).
In total there were 15 Galactic source scans:
one of the Carina Nebula, four of RCW38, four of IRAS1022, five of IRAS08576
and one which included both NGC3603 and NGC3576. Nearly (but not quite)
all of the data for all of the detectors was good for each scan;
Table \ref{tbl1} lists the number of good observations
(which includes the number of good scans and channels) at each frequency.
Scans of the Galactic sources were flagged as bad for a particular
channel for example if the source was on the edge of the scan in that channel.
The extra-galactic sources were observed serendipitously during
the CMB scans during the entire flight. The total number
of good observations of the extra-galactic sources in
Table \ref{tbl1} reflects the number of channels
at each frequency in which the source was detected.

\begin{figure}
\epsscale{0.75}
\plotone{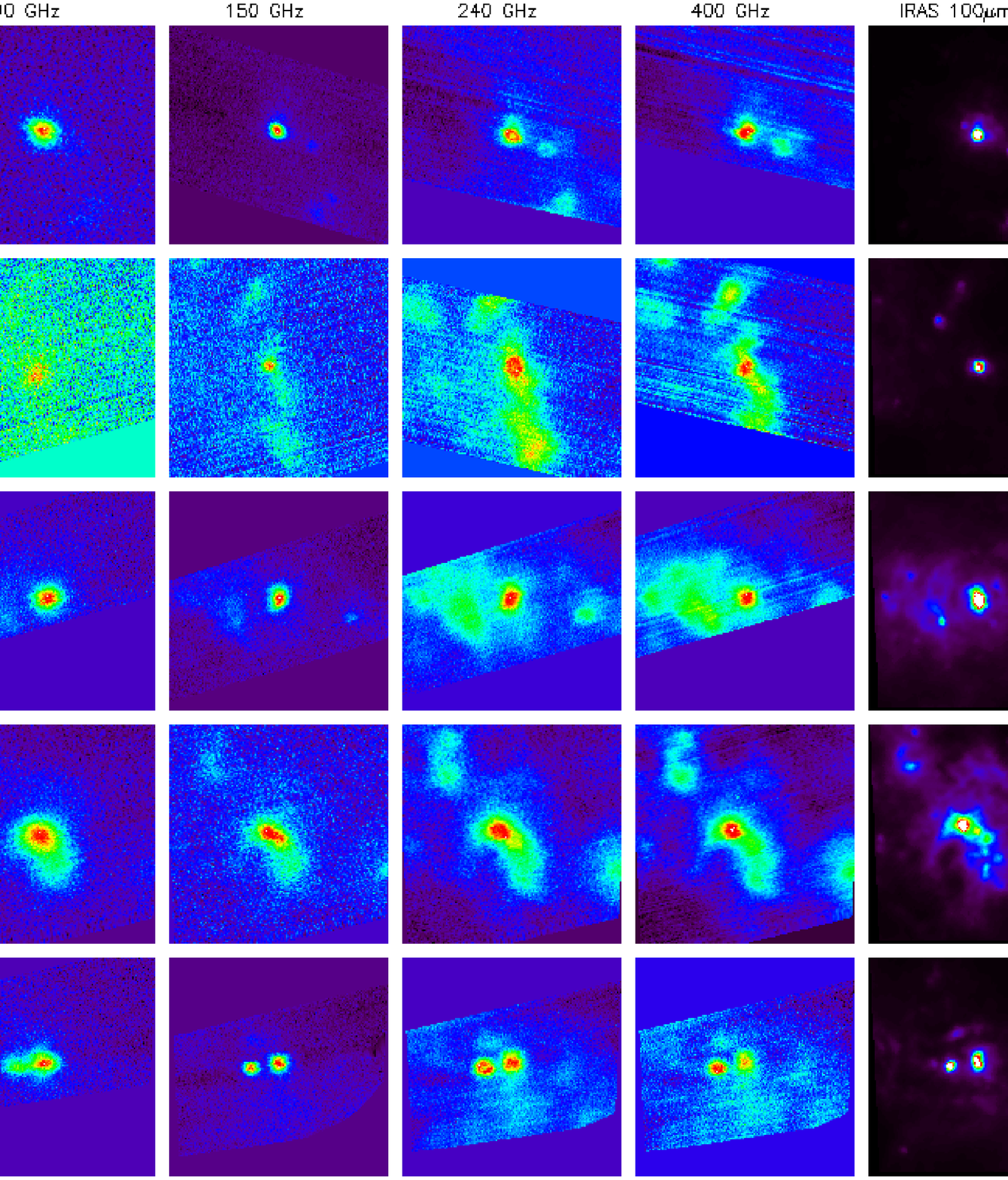}
\caption{BOOMERANG maps of the Galactic sources, made from data from a single scan and a single channel. Other observations for each source and frequency are similar and the total number of observations for each source at each frequency are given in Table \ref{tbl1}. IRAS 100 $\mu$m maps are included for comparison. The images are all $3^{\circ} \times 3^{\circ}$ in extent. The maps are flat space projections centered on the source, with coordinates given by ($x=(\alpha-\alpha_0)cos(\delta)$, $y=(\delta-\delta_0$)), where ($\alpha_0$, $\delta_0$) are the Right Ascension and declination of the sources given in Table \ref{tbl1}.}
\label{fig1}
\end{figure}

BOOMERANG maps of the Galactic sources are shown in Figure
\ref{fig1}. Each of the maps shown are 3 degrees $\times$ 3 degrees
in extent and represent only one channel of data for one scan
(approximately 15 minutes of data). Other scans
are similar and the total number of observations for each
source at each frequency are given in Table \ref{tbl1}
The maps are flat space projections centered on the source, with
coordinates given by ($x=(\alpha-\alpha_0)cos(\delta)$, $y=(\delta-\delta_0$)),
where ($\alpha_0$, $\delta_0$) are the Right Ascension and declination
of the sources given in Table \ref{tbl1}.
IRAS 100 $\mu m$ maps (data obtained through the Skyview website)
of the same regions are included for comparison. 
From the Galactic source maps, NGC3576 is the most compact of our
sources, followed by RCW38. IRAS1022 and especially the Carina Nebula
show significant extended emission. IRAS08576 shows significant extended
emission along a line; this is unexpected from IRAS 100 $\mu m$ data,
which shows IRAS08576 as a compact object. In the maps there are
bright, compact sources typically surrounded by more diffuse clouds.
Fluxes are measured for the bright, compact sources in each map. 

\begin{figure}
\epsscale{0.75}
\plotone{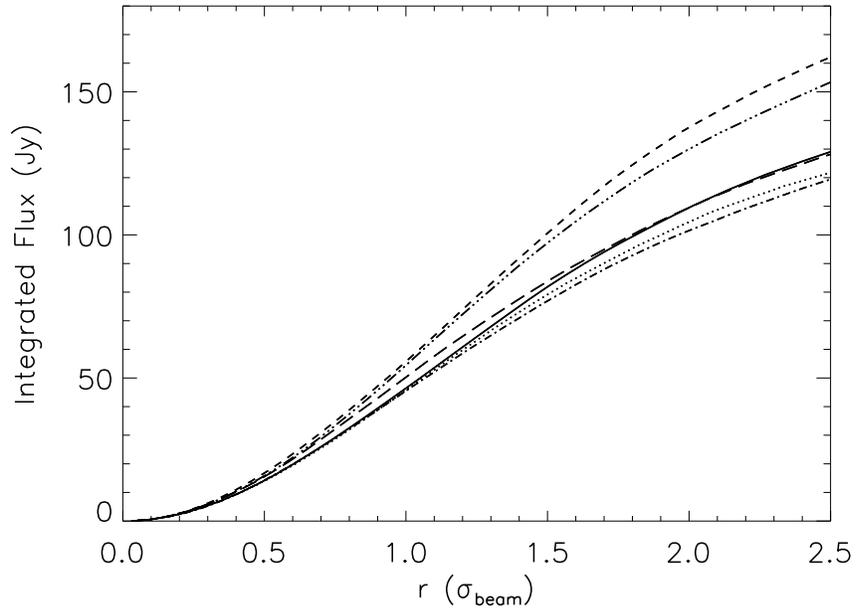}
\caption{\footnotesize Integrated flux vs. radius out from the center of the source for one of the 150 GHz scans of RCW38. The six different curves are for the six 150 GHz channels. Radius is given in units of the beamsize $\sigma_{beam}$ because the beamsize is slightly different for each channel.}
\label{fig2}
\end{figure}

The Galactic source maps for each scan and channel
are binned in radius around the
centroid of the source to obtain the integrated flux
as a function of radius. A sample radial flux profile
for one source and frequency is shown in Figure \ref{fig2}.
Following the notation of Puchalla et al. (2002), we calculate
the total integrated flux within radii of 1 and 2 $\sigma_{beam}$
(hereafter 1$\sigma_{b}$ and 2$\sigma_{b}$ fluxes). 
The beam size is given by $\sigma_{beam}=FWHM/(2\sqrt{2ln2})$ and
the FWHMs of the detectors are given above.
When computing fluxes for NGC3603 and NGC3576, the
companion source is masked out in the map.

Table \ref{tbl2} lists the 1$\sigma_{b}$ and 2$\sigma_{b}$ fluxes
for all sources and frequencies.
The flux and uncertainty for the Galactic sources
in Table \ref{tbl2} are the mean and standard
deviation of all of the good channels and scans at the given frequency.
Only the statistical uncertainty is given in the table. 
Uncertainty in the flux due to calibration is
not included. The $1 \sigma_{b}$ and $2 \sigma_{b}$ flux results
are systematically dependent on beam
at approximately the 10\% and 5\% level, respectively. The 
fluxes are dependent on background subtraction at about the 1\% level.
The uncertainty in the flux due to the inherent S/N of the
data is approximately 2\%.

Since the S/N on the extra-galactic sources is much lower
than the S/N on the Galactic sources, maps are not used in
this analysis;
fluxes are computed directly by binning the timestream data in radius,
after filtering at 0.2 Hz. Since there are relatively
few observations of each source (see Table \ref{tbl1})
and since the S/N of each observation is low, {\em weighted} means
and standard deviations are given in Table \ref{tbl2}.
Background subtraction systematically affects the 1$\sigma_{b}$
and 2$\sigma_{b}$ fluxes by about 10\% and 25\% respectively.
Uncertainty in the beam systematically affects the
1$\sigma_{b}$ and 2$\sigma_{b}$ fluxes by approximately
30\% and 25\% respectively.

If the flux of the source scales as $f \propto \nu^{\alpha}$,
then the spectral index, $\alpha_{1:2}$, between frequencies
$\nu_1$ and $\nu_2$, is given by:
\begin{equation}
\alpha=\frac{log(f_2/f_1)}{log(\nu_2/\nu_1)}.
\label{eq:specindex}
\end{equation}
Spectral indices for the BOOMERANG 1$\sigma_{b}$ and 2$\sigma_{b}$ fluxes
are given in Table \ref{tbl3}.

\placetable{tbl2}

\placetable{tbl3}


\section{Wavelet Analysis of the Extra-galactic Sources}

In addition to the analysis in the previous section,
a wavelet approach is also used, following Vielva et al. (2001),
to evaluate the fluxes and the spectral indices 
for the three extra-galactic point sources observed by BOOMERANG.
The main advantages of using this method are the amplification of
the signal to noise ratio and that the result is only sensitive to
sources of size equal to or smaller than the beam, so
the measured fluxes are much less affected by the background signal.

The isotropic wavelet transform for a two-dimensional map 
$T(\mathbf{x})$ is defined by:
\begin{equation}\label{eq:wav}
        w_T (R, \mathbf{p}) = \int d^2 \mathbf{x} \frac{1}{R^2}
                \psi \left( \frac{|\mathbf{x}-\mathbf{p}|}{R} \right)
                T(\mathbf{x})
\end{equation}
where $w_T (R, \mathbf{p})$ is the wavelet coefficient of the map $T$
associated with a scale $R$ and centered at the point $\mathbf{p}$.
The function $\psi (|\mathbf{x}|)$ is called the wavelet mother function.
If the beam profile is Gaussian it is found that Mexican 
Hat wavelet is the optimal one. The mother function is then
\begin{equation}\label{eq:mother}
        \psi (x) = \frac{1}{\sqrt{2 \pi}}\left[ 2-
                \left(\frac{2}{R}\right)^2 \right] 
                e^{\frac{x^2}{2 R^2}}
\end{equation}
and has zero integral.
Substituting eq. \ref{eq:mother} into eq. \ref{eq:wav},
for a point source centered on $\mathbf{y}$ yields
\begin{equation}\label{eq:th}
        w_T^{th}(R, \mathbf{y}) = 2 \sqrt{2\pi} \frac{A}{\Omega}
                \frac{(R/\sigma_b)^2}{(1+(R/\sigma_b)^2)^2}
\end{equation}
where $\Omega$ is the area under the Gaussian beam, $\sigma_b$
is the beam dispersion and $A$ is the flux of the point source.
In order to estimate the point source flux, $A$, a multi-scale
fit is performed computing
the wavelet coefficients of our map, centered in the nominal 
coordinates of the source, at a number of scales $R$.
The results are compared with the theoretical curve given by 
eq. \ref{eq:th} using the $\chi^2$ statistic:
\begin{equation}
\chi^2 = \left[ w_T(R_i, \mathbf{y}) - w_T^{th}(R_i, \mathbf{y}) \right]
        C^{-1}(R_i, R_j)
        \left[ w_T(R_j, \mathbf{y}) - w_T^{th}(R_j, \mathbf{y}) \right]
\end{equation}
where the covariance matrix $C$ is computed using real data:
\begin{equation}
        C_{i j} = \langle w_T(R_i,\mathbf{x})w_T(R_j,\mathbf{x})\rangle
\end{equation}
and the average is performed over random positions $\mathbf{x}$.

In order to check the effect of a non-Gaussian beam on the results,
the analysis has been done on simulated point sources with different
beam shapes. The error in the flux recovery is smaller than $10\%$
even for a highly non-Gaussian beam, such as a square or triangular beam,
if there is a good estimation of the integrated 
beam, $\Omega$, in eq. \ref{eq:th}.

Spectral indices are computed from the fluxes according
to eq. \ref{eq:specindex}. The flux and spectral index
results are given in Table \ref{tbl4} and are in good
agreement with the radial integration results of the previous section.

\placetable{tbl4}


\section{SEST Observations and Results}

Point source observations of the three extra-galactic sources and
images of the Galactic sources RCW38 and NCG3576 at 90 and 150GHz
were made with the SEST telescope at La Silla Observatory in Chile
on January 8-10, 2000. The FWHM of the SEST 90 GHz beam is 57 arcsec,
the FWHM of the SEST 150 GHz beam is 35 arcsec and the chop is 11 arcmin.
Calibration was done using the chopper wheel method. The antenna
temperature scale, given in $T_{A*}$, was converted to a flux density
scale through observations of planets (SEST web page).

For the Galactic sources NGC3576 and RCW38, 
observations were made in a region 4'x4' in extent
centered on each source with a grid spacing of 17.5 arcsec
(hereafter center maps).
Additional observations were made with a
wider grid spacing (35 arcsec) covering a larger region
around each source (10 arcmin x 10 arcmin in extent for
NGC3576, 6 arcmin x 6 arcmin in extent for RCW38) (hereafter extended maps).
For RCW38, the more extended of the two sources,
the center and extended observations were combined.
For NCG3576 only the center maps were used, since the
data far from the source are noisy or could be contaminated
by objects outside the field.

\begin{figure}
\epsscale{0.75}
\plotone{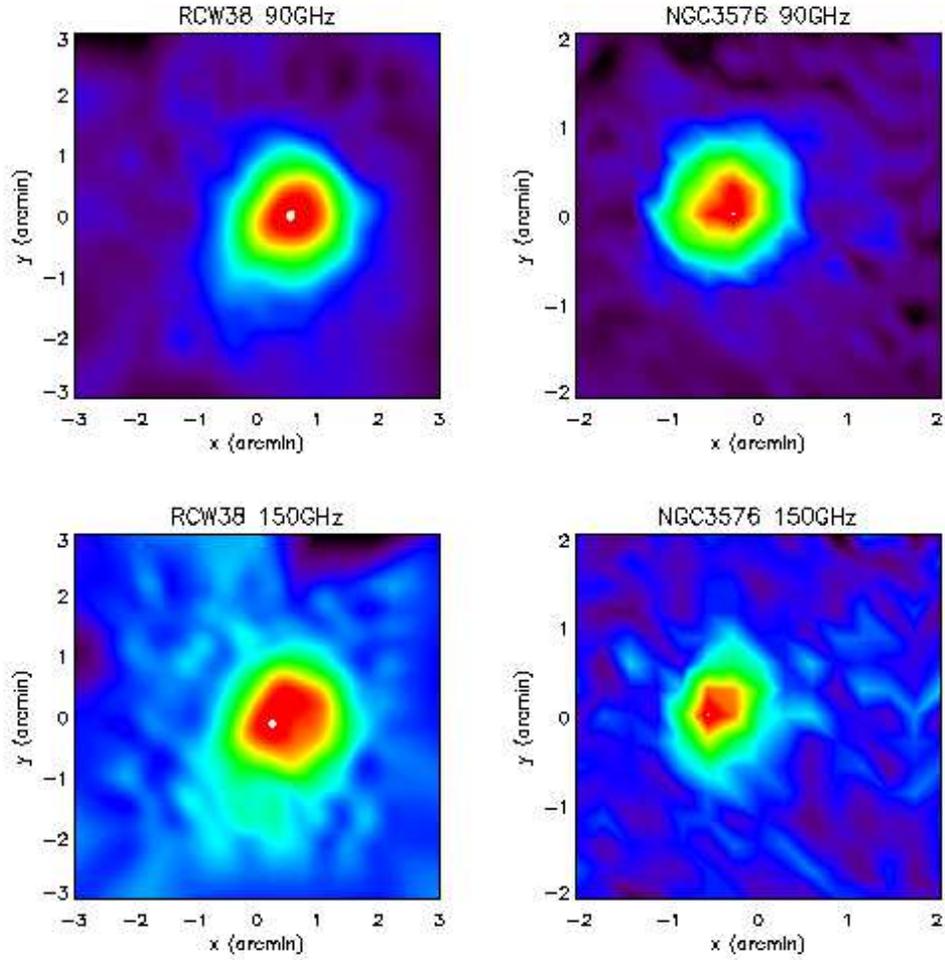}
\caption{SEST maps of compact Galactic sources RCW38 and NGC3576. The maps are flat space projections centered on the source, with coordinates given by ($x=(\alpha-\alpha_0)cos(\delta)$, $y=(\delta-\delta_0$)), where ($\alpha_0$, $\delta_0$) are the Right Ascension and declination of the sources given in Table \ref{tbl1}.}
\label{fig3}
\end{figure}

Maps are shown in Figure \ref{fig3} and fluxes are given
in Table \ref{tbl5}. Again, maps are a flat-space
projection centered on the source, as in Figure \ref{fig1}.
Fluxes for RCW38 and NGC3576 are
computed within the map regions. Fluxes for the extra-galactic sources
are from point source observations. Spectral indices are
computed from the fluxes using eq. \ref{eq:specindex} and are
also given in Table \ref{tbl5}.

The SEST observations can be convolved with the BOOMERANG beam in order to
check the BOOMERANG calibration. 
For RCW38 and NGC3576, the SEST maps are convolved with
the BOOMERANG beam and flux profiles are obtained from the convolved maps.
A numerical convolution is not necessary for the extra-galactic sources 
since the analytic convolution of a point source with a Gaussian is
simple; the 1 $\sigma_{b}$ and 2 $\sigma_{b}$ fluxes should be 0.393 and 0.865
of the point source flux. Flux ratios are given in Table \ref{tbl6},
where the BOOMERANG fluxes used are those from Table \ref{tbl2}.
The extra-galactic sources may be time variable, and
since the BOOMERANG and SEST observations were made approximately 
a year apart, they could have different intensities not only because of 
calibration but because of intensity variations.
From a weighted mean of the numbers in Table \ref{tbl6},
the current BOOMERANG calibration relative to SEST
is 0.97$\pm$0.07 for the 90 GHz channels
and 1.04$\pm$0.06 for the 150 GHz channels. In other words, the BOOMERANG
and SEST calibrations agree with each other to within the uncertainties.

The MAT team has published fluxes for NGC3576 at 150GHz (Puchalla et al. 2002).
For MAT, $\sigma_{beam} = 12.14/(2\sqrt{2ln2})$ = 5.16 arcmin. 
The ratio MAT/(SEST*MAT) for the NGC3576 150 GHz
observations is 1.01$\pm$0.14 and 1.00$\pm$0.15 for their
$1\sigma_{b}$ and $2\sigma_{b}$ fluxes respectively,
implying that SEST and MAT agree well.
Therefore, the calibrations of BOOMERANG, SEST and MAT are
all consistent for NGC3576 at 150 GHz.

\placetable{tbl5}

\placetable{tbl6}


\section{Conclusions}

A comparison with the SEST data for the Galactic sources
NGC3576, RCW38 and extra-galactic sources 0537-441, 0521-365 and
0438-43 implies that the current BOOMERANG calibration and
SEST agree to within 8\% at 90 GHz and 6\% at 150 GHz.
Additionally, BOOMERANG, MAT, and SEST fluxes of NGC3576 at 150 GHz
are all consistent with each other.

\begin{figure}
\epsscale{0.75}
\plotone{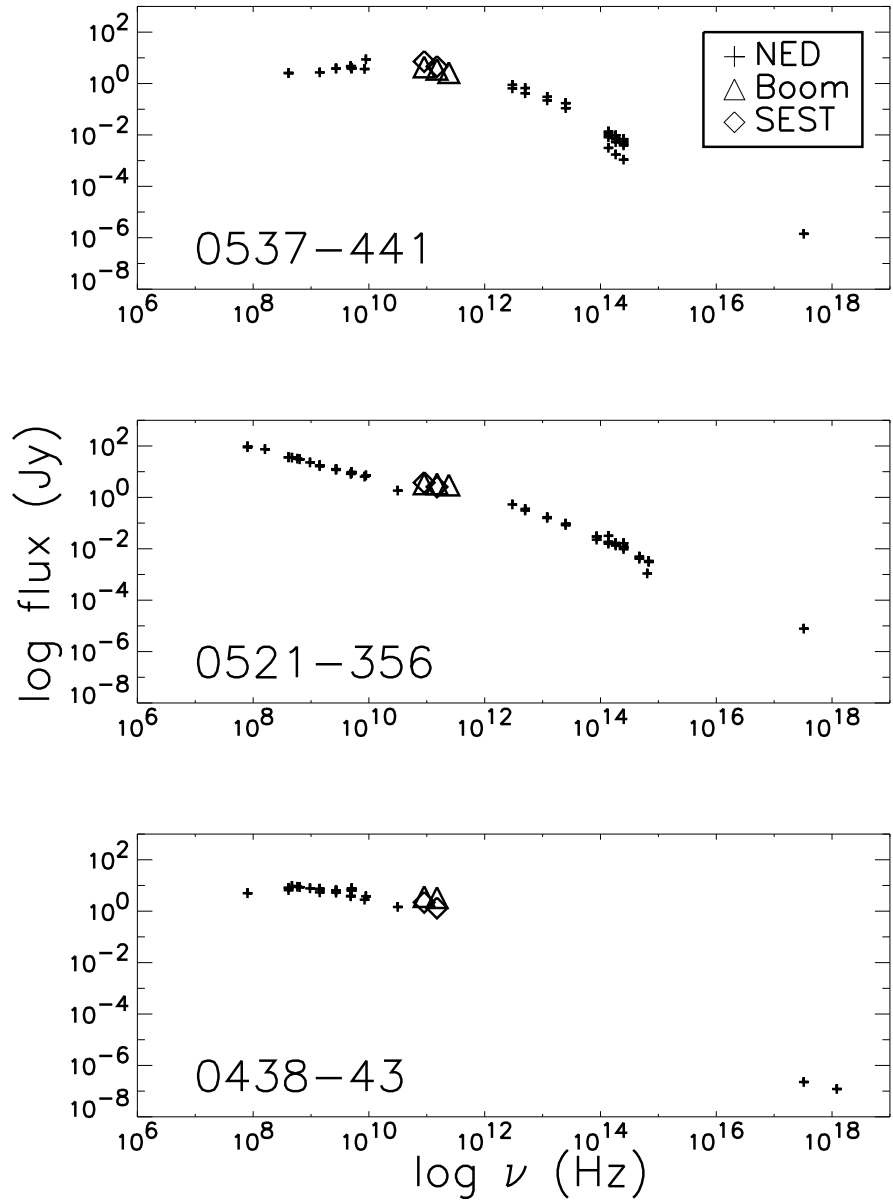}
\caption{Fluxes of extra-galactic sources.
BOOMERANG and SEST fluxes for each source are plotted along with measurements
over a wide range of the electromagnetic spectrum from the NASA/IPAC
Extra-galactic Database (NED).}
\label{fig4}
\end{figure}

The Galactic sources typically show a flat to inverted spectrum from 90 to
150 GHz and a rising spectrum from 150 to 240 GHz and from
240 to 400 GHz. This is consistent with a combination of free-free
and dust emission, with free-free dominating at the lower
frequencies and dust dominating at higher frequencies.
Our spectral indices are consistent with the typical
$f \propto \nu^2$ expected for millimeter observations of
HII regions. While the uncertainties on the spectral
indices of the extra-galactic sources from BOOMERANG
are large, the SEST measurements imply a falling spectrum
with increasing frequency. To place these measurements in context,
Figure \ref{fig4} shows the extra-galactic source fluxes
along with measurements from a wide range of the electromagnetic spectrum.


{\it Acknowledgments}:

We gratefully thank the night assistants at SEST for helping us
with the observations. We thank Paula Ehlers for valuable input.
The BOOMERANG program has been supported by NASA (NAG5-4081 \& NAG5-4455),
the NSF Science \& Technology Center for Particle 
Astrophysics (SA1477-22311NM under AST-9120005) and
NSF Office of Polar Programs (OPP-9729121) in the USA,
Programma Nazionale Ricerche in Antartide, Agenzia Spaziale Italiana and 
University of Rome La Sapienza in Italy, and by PPARC in UK.
The Swedish-ESO Submillimetre Telescope, SEST, is operated jointly 
by ESO and the Swedish National Facility for Radioastronomy,
Onsala Space Observatory at Chalmers University of Technology.
KC is supported by NSF grant AST-0104465 under the AAPF program.



\clearpage

\clearpage

\clearpage

\clearpage

\clearpage

\begin{table*}
\begin{center}
\scriptsize
\begin{tabular}{|cccccccc|}
\tableline
Source		&RA	&Dec 	&Object Type	&$N_{90}$	&$N_{150}$	&$N_{240}$	&$N_{400}$\\
\tableline
Carina		&161.06	&-59.58	&HII Region	&2		&6		&3		&4\\
RCW38		&134.77	&-47.51	&HII Region	&8		&24		&12		&16\\
IRAS1022	&156.05	&-57.80	&HII Region	&7		&22		&11		&13\\
NGC3603		&168.79	&-61.29	&HII Region	&2		&6		&3		&4\\
NGC3576		&167.98	&-61.33	&HII Region	&2		&6		&3		&4\\
IRAS08576	&134.86	&-43.76	&HII Region	&10		&29		&15		&19\\
0537-441	&84.71	&-44.09	&Blazar		&2		&6		&1		&0\\
0521-365	&80.74	&-36.46	&BL Lac		&2		&5		&2		&0\\
0438-43		&70.07	&-43.55	&QSO		&1		&4		&0		&0\\
\tableline
\end{tabular}
\end{center}
\caption{Observed sources. Listed are J2000 coordinates, object type and number of observations by BOOMERANG at each frequency. The total number of observations at each frequency (e.g. $N_{90}$ at 90~GHz) is the number of good scans over all channels at that frequency. Coordinates of Galactic sources are from IRAS maps. Coordinates of extra-galactic sources are from SEST.}
\label{tbl1}
\end{table*}

\clearpage

\begin{table*}
\begin{center}
\scriptsize
\begin{tabular}{|c|cc|cc|cc|cc|}
\tableline
Source		&90 GHz		&90 GHz		&150 GHz	&150 GHz	&240 GHz	&240 GHz	&400 GHz	&400 GHz\\
		&1$\sigma_{b}$ (Jy)	&2$\sigma_{b}$ (Jy)	&1$\sigma_{b}$ (Jy)	&2$\sigma_{b}$ (Jy)	&1$\sigma_{b}$ (Jy)	&2$\sigma_{b}$ (Jy)	&1$\sigma_{b}$ (Jy)	&2$\sigma_{b}$ (Jy)\\
\tableline
Carina		&72.4$\pm$8.2	&214$\pm19$	&37.9$\pm$4.1	&131$\pm$13	&132$\pm$18	&410$\pm$56	&446$\pm$145	&1324$\pm$419\\
RCW38		&60.8$\pm$6.2	&137$\pm$16	&51.7$\pm$4.0	&122$\pm$12	&126$\pm$17	&298$\pm$42	&406$\pm$115	&1002$\pm$298\\
IRAS1022	&60.7$\pm$4.6	&143$\pm$15	&42.2$\pm$2.9	&119$\pm$11	&107$\pm$15	&282$\pm$38	&320$\pm$94	&904$\pm$278\\
NGC3603		&55.6$\pm$6.1	&126$\pm$14	&42$\pm$1.2	&107$\pm$4.4	&101$\pm$14	&249$\pm$35	&308$\pm$99	&802$\pm$256\\
NGC3576		&32.6$\pm$3.6	&74.6$\pm$9.1	&32.7$\pm$1.5	&70.1$\pm$6.0	&87.9$\pm$10	&201$\pm$24	&360$\pm$102	&727$\pm$196\\
IRAS08576	&9.1$\pm$0.7	&20.6$\pm$2.6	&11.8$\pm$0.7	&31.9$\pm$2.2	&55.2$\pm$6.3	&158$\pm$18	&230$\pm$59	&667$\pm$173\\
0537-441	&1.8$\pm$0.4	&3.8$\pm$1.1	&1.5$\pm$0.2	&3.1$\pm$0.6	&1.3$\pm$0.8	&2.3$\pm$2.1	&--	&--\\
0521-365	&1.5$\pm$0.5	&2.9$\pm$1.4	&1.3$\pm$0.3	&2.9$\pm$0.7	&1.3$\pm$0.7	&2.6$\pm$0.4	&--	&--\\
0438-43		&1.4$\pm$0.5	&3.2$\pm$1.3	&1.2$\pm$0.2	&2.9$\pm$0.6	&--		&--		&--	&--\\
\tableline
\end{tabular}
\end{center}
\caption{Fluxes in Jy of sources as determined by observations with BOOMERANG. The 1$\sigma_{b}$ and 2$\sigma_{b}$ fluxes are the integrated fluxes within radii of 1 and 2 $\sigma_{beam}$ respectively. Errors are statistical only; they do not include systematic errors due to uncertainty in the beam size, background subtraction or calibration. For a point source observed with a gaussian beam,
the integrated fluxes out to 1$\sigma_{b}$ and 2$\sigma_{b}$
are 0.393 and 0.865 of the source flux, respectively.
The extra-galactic sources are point sources,
whereas the Galactic sources are somewhat extended.}
\label{tbl2}
\end{table*}

\clearpage

\begin{table*}
\begin{center}
\scriptsize
\begin{tabular}{|c|cc|cc|cc|}
\tableline
Source	&$\alpha_{90:150}$&$\alpha_{90:150}$&$\alpha_{150:240}$&$\alpha_{150:240}$&$\alpha_{240:400}$&$\alpha_{240:400}$\\
		&1$\sigma_{b}$	&2$\sigma_{b}$	&1$\sigma_{b}$	&2$\sigma_{b}$	&1$\sigma_{b}$	&2$\sigma_{b}$	\\
\tableline
Carina		&-1.3$\pm$0.3	&-1.0$\pm$0.3	&3.3$\pm$0.5	&3.0$\pm$0.5	&2.0$\pm$0.6	&2.0$\pm$0.6	\\
RCW38		&-0.3$\pm$0.3	&-0.2$\pm$0.3	&2.3$\pm$0.5	&2.3$\pm$0.5	&2.0$\pm$0.6	&2.0$\pm$0.6	\\
IRAS1022	&-0.7$\pm$0.2	&-0.4$\pm$0.3	&2.4$\pm$0.4	&2.3$\pm$0.5	&1.8$\pm$0.6	&1.9$\pm$0.6	\\
NGC3603		&-0.6$\pm$0.3	&-0.3$\pm$0.3	&2.3$\pm$0.4	&2.2$\pm$0.4	&1.9$\pm$0.6	&2.0$\pm$0.6	\\
NGC3576		&0.0$\pm$0.3	&-0.1$\pm$0.3	&2.6$\pm$0.4	&2.7$\pm$0.4	&2.4$\pm$0.5	&2.2$\pm$0.5	\\
IRAS08576	&0.5$\pm$0.2	&0.9$\pm$0.3	&4.0$\pm$0.4	&4.2$\pm$0.4	&2.4$\pm$0.5	&2.4$\pm$0.5	\\
0537-441	&-0.4$\pm$0.5	&-0.5$\pm$0.7	&-0.4$\pm$1.6	&-0.7$\pm$2.5	&--		&--		\\
0521-365	&-0.3$\pm$0.8	&0.0$\pm$1.0	&0.1$\pm$1.5	&-0.1$\pm$2.0	&--		&--		\\
0438-43		&-0.4$\pm$0.8	&-0.2$\pm$0.9	&--		&--		&--		&--		\\
\tableline
\end{tabular}
\end{center}
\caption{Spectral indices of sources observed with BOOMERANG. The spectral index $\alpha_{f1:f2}$ between two frequencies $f1$ and $f2$ is computed using eq. \ref{eq:specindex} and the fluxes in Table \ref{tbl2}.}
\label{tbl3}
\end{table*}

\clearpage

\begin{table*}
\begin{center}
\scriptsize
\begin{tabular}{|c|c|c|c|c|c|}
\tableline
Source   &   90 GHz (Jy)  &  150 GHz (Jy)  &  240 GHz (Jy)  & $\alpha_{90:150}$ & $\alpha_{150:240}$ \\
\tableline
0537-441 & $3.8\pm0.8$ & $3.6\pm0.6$ & $2.2\pm0.4$ & $-0.1\pm0.5$ & $-0.9\pm0.6$ \\
0521-365 & $3.4\pm0.8$ & $2.8\pm0.4$ & $2.7\pm0.5$ & $-0.3\pm0.5$ & $-0.1\pm0.6$ \\
0438-43 & $2.1\pm0.5$ & $1.5\pm0.3$ & $1.0\pm0.2$ & $-0.6\pm0.5$ & $-0.8\pm0.7$ \\
\tableline
\end{tabular}
\end{center}
\caption{Total fluxes in Jy and spectral indices of extra-galactic
sources observed by BOOMERANG computed using a wavelet analysis.}
\label{tbl4}
\end{table*}

\clearpage

\begin{table*}
\begin{center}
\scriptsize
\begin{tabular}{|c|c|c|c|}
\tableline
Source		&90 GHz (Jy)	&150 GHz (Jy)	&$\alpha_{90:150}$\\
\tableline
RCW38		&125$\pm$10	&116$\pm$7	&-0.15$\pm$0.20	\\
NGC3576		&75$\pm$7	&78$\pm$8	&0.08$\pm$0.27	\\
0537-441	&7.21$\pm$0.31	&4.59$\pm$0.35	&-0.9$\pm$0.2	\\
0521-365	&3.76$\pm$0.30	&2.55$\pm$0.41	&-0.8$\pm$0.4	\\
0438-43		&2.21$\pm$0.31	&1.33$\pm$0.41	&-1.0$\pm$0.7	\\
\tableline
\end{tabular}
\end{center}
\caption{Fluxes in Jy and spectral indices of sources observed with SEST. Fluxes for RCW38 and NGC3576 are computed for the regions shown in Figure \ref{fig3}. Fluxes for the extra-galactic sources are from point source observations.}
\label{tbl5}
\end{table*}

\clearpage

\begin{table*}
\begin{center}
\scriptsize
\begin{tabular}{|c|cc|cc|}
\tableline
Source		&B/S 90 GHz	&B/S 90 GHz	&B/S 150 GHz	&B/S 150 GHz	\\
		&1$\sigma_{b}$	&2$\sigma_{b}$	&1$\sigma_{b}$	&2$\sigma_{b}$	\\
\tableline
RCW38		&1.25$\pm$0.17	&1.27$\pm$0.19	&1.15$\pm$0.12	&1.22$\pm$0.15	\\
NGC3576		&1.11$\pm$0.16	&1.15$\pm$0.17	&1.09$\pm$0.12	&1.05$\pm$0.15	\\
0537-441	&0.65$\pm$0.14	&0.62$\pm$0.18	&0.85$\pm$0.13	&0.77$\pm$0.15	\\
0521-365	&1.02$\pm$0.34	&0.87$\pm$0.42	&1.27$\pm$0.30	&1.33$\pm$0.37	\\
0438-43		&1.61$\pm$0.60	&1.69$\pm$0.73	&2.20$\pm$0.79	&1.95$\pm$0.72	\\
\tableline
\end{tabular}
\end{center}
\caption{Flux ratios of BOOMERANG to the convolution of the SEST data with the BOOMERANG beam: BOOM/(SEST*BOOM). Errors are statistical only; they do not include the systematic error in the fluxes due to uncertainty in the BOOMERANG beam size. The BOOMERANG fluxes used are those from Table \ref{tbl2}.
The extra-galactic sources may be time variable, and
since the BOOMERANG and SEST observations were made approximately 
a year apart, they could have different intensities not only because of 
calibration but because of intensity variations.}
\label{tbl6}
\end{table*}


\end{document}